# Carbon nanotube substrates enhance SARS-CoV-2 spike protein ion yields in matrix assisted laser desorption-ionization mass spectrometry

T. Schenkel[1], A. M. Snijders[2], K. Nakamura[1], P.A. Seidl[1], B. Mak[1], L. Obst-Huebl[1], H. Knobel[3], I. Pong[1], A. Persaud[1], J. van Tilborg[1], T. Ostermayr[1], S. Steinke[1], E. A. Blakely[1], Q. Ji[1], A. Javey[4], R. Kapadia[5], C.G.R. Geddes[1], E. Esarey[1]

[1]Accelerator Technology and Applied Physics Division, Lawrence Berkeley National Laboratory, Berkeley, CA 94720, USA
[2] Biological Systems and Engineering Division, Lawrence Berkeley National Laboratory, Berkeley, CA 94720, USA
[3]Eurofins Materials Science Netherlands BV, 5656 AE Eindhoven, The Netherlands
[4]Electrical Engineering and Computer Science Department, University of California, Berkeley, CA 94720, USA
[5]School of Engineering, University of Southern California, Los Angeles, CA 90089, USA

Corresponding author: T_Schenkel@LBL.gov

Nanostructured surfaces enhance ion yields in matrix assisted laser desorption-ionization mass spectrometry (MALDI-MS). The spike protein complex, S1, is one fingerprint signature of Sars-CoV-2 with a mass of 75 kDa. Here, we show that MALDI-MS yields of Sars-CoV-2 spike protein ions in the 100 kDa range are enhanced 50-fold when the matrix-analyte solution is placed on substrates that are coated with a dense forest of multi-walled carbon nanotubes, compared to yields from uncoated substrates. Nanostructured substrates can support the development of mass spectrometry techniques for sensitive pathogen detection and environmental monitoring.

The global SARS-CoV-2 pandemic since early 2020 has highlighted the need for low cost, sensitive, robust and widely deployable techniques for detection of pathogens, such as the SARS-CoV-2 virus. Mass spectrometry techniques are very widely used in bio-medical sciences. MALDI-MS (matrix-assisted laser desorption/ionization mass spectrometry) is a relatively low cost and rapid technique [1]. In response to the current crisis, MALDI-MS based detection of SARS-CoV-2 signatures has been reported from nasal swabs and from saliva samples [2-5]. In MALDI-MS, the use of (nano-)structured substrates has been shown to enhance analyte ion yields and, in some cases, eliminate the need for a matrix altogether [7-10]. However, these signal increases were limited to relatively low mass analyte ions, below 10 kDa. Here, we report on the detection of SARS-CoV-2 spike protein (S1) ions in the 100 kDa mass range. The spike protein complex is a signature for detection of SARS-CoV-2 [2-4]. With a mass of 75 kDa, direct detection in MALDI-MS is challenging as the desorption and ionization process in the laser-matrix-analyte interaction can lead to fragmentation of the protein into a series of poly-peptide and small fragment ions. This makes unique identification of pathogens difficult in MALDI-MS experiments. Detection of robust signatures in mass spectra of SARS-CoV-2 samples is beneficial for the development of environmental sensors and for rapid, low-cost testing capabilities outside laboratory settings. We compare spike protein ion yields from carbon nanotube substrates to yields from standard flat substrates (made from single crystal silicon wafers) for a series of MALDI-MS conditions and find S1 analyte ion yields that are enhanced 50-fold when we used carbon nanotube substrates.

Recombinant SARS-CoV-2 spike protein, S1, was purchased from R&D Systems (Minneapolis, MN, USA). We prepared dilute solutions of S1 with sinapinic acid as a standard MALDI-MS matrix (purchased from ProteoChem, Hurricane, UT, USA). The matrix analyte solution consisted of 10 mg of sinapinic acid in a 1 ml of solution (50% acetonitrile and 50% water with 0.1% trifluoroacetic acid) and we prepared analyte – matrix samples with 0.05 to 0.2 µg of S1 protein per 1 µl of matrix solution. We then deposited drops of 2.5 µl of matrix-analyte solution onto flat silicon single crystal samples and onto silicon samples which had been coated with a dense forest of multi-wall carbon nanotubes (CNT).

    Aligned CNT were grown on silicon wafers in a plasma-enhanced chemical vapor deposition (PECVD) process [11]. The growth process results in a relatively uniform height of CNT. A scanning electron microscope (SEM) image of a sample with a CNT forest is shown in Figure 1 (a). The average diameter of the CNTs is about 70 nm. In Figure 1 (b) we show a section of a dried drop of matrix-analyte solution deposited on a CNT substrate. SEM images of mm-scale dried drops of matrix-analyte solution on CNT and flat substrates are shown in Fig 1 (c, d, respectively). We observe formation of much denser assemblies of matrix-analyte crystals on the CNT samples compared to flat substrates (silicon shown, also titanium, not shown). The images (Fig 1, bottom row) also show damage spots from sample exposure to laser pulses during MALDI-MS experiments.

a)

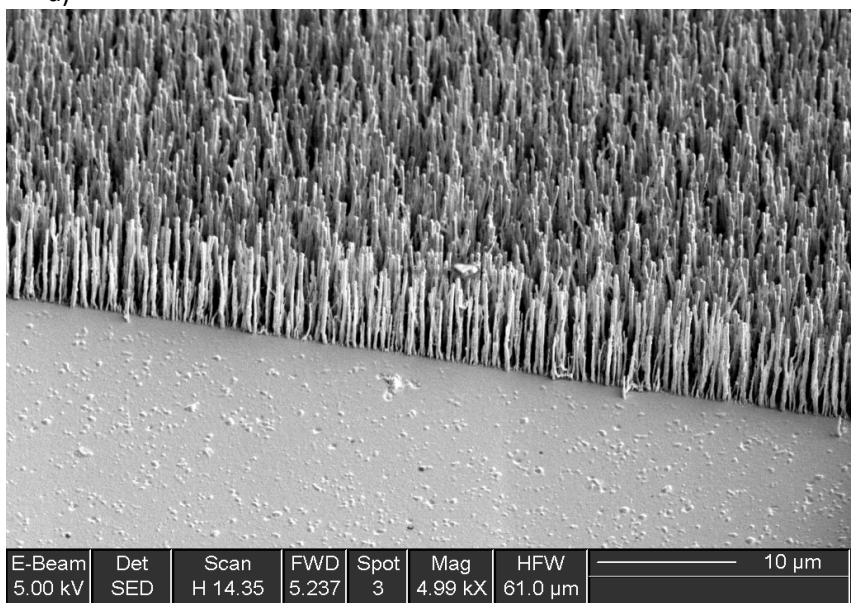

b)

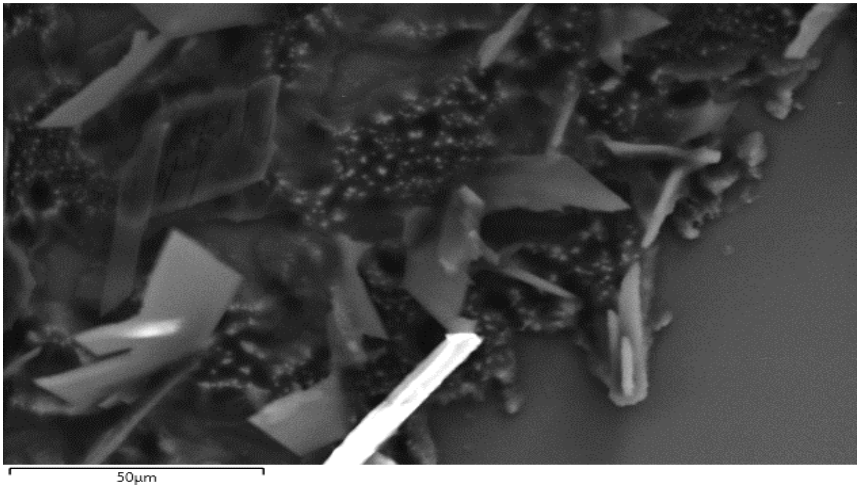

c)

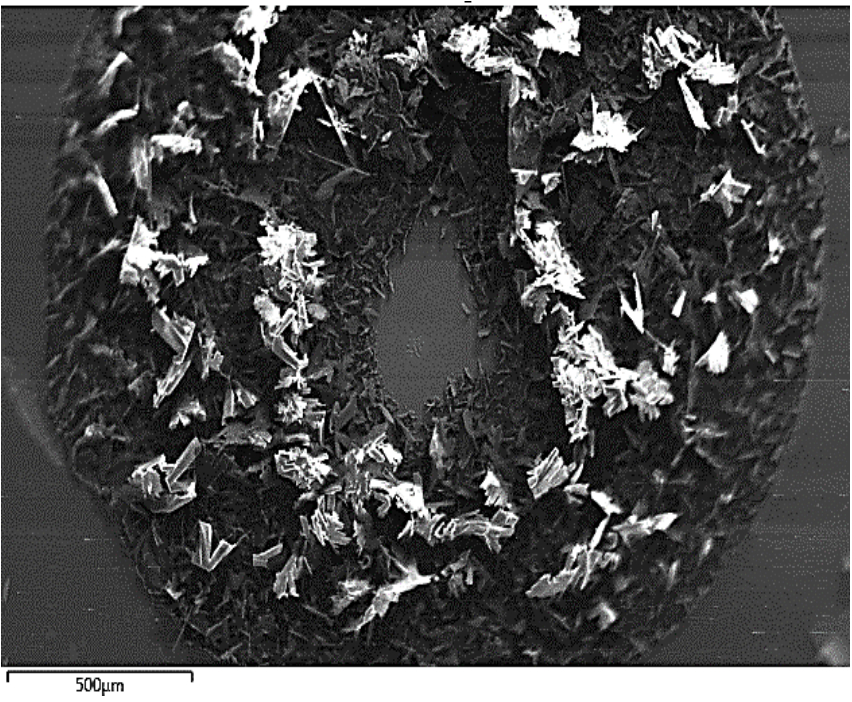

d)

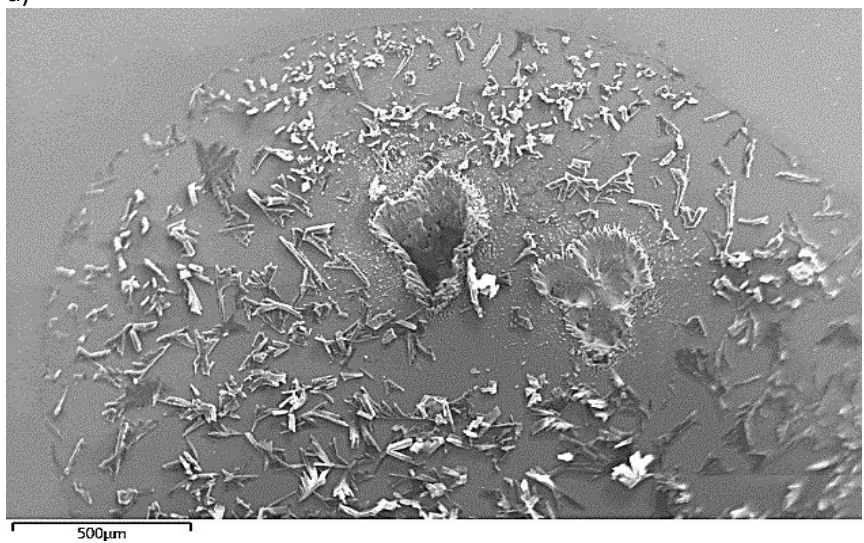

Figure 1: SEM images of a CNT sample without (a) and with MALDI matrix crystals (b). c) SEM images of matrix-analyte crystals on a substrate with CNTs and on a flat silicon single crystal substrate (d) after exposure to laser pulses in MALDI experiments showing local depletion of matrix crystals (c) and ablation of the silicon substrate (d).

We conducted MALDI-MS experiments with the Sciex 4800 MALDI-TOF-TOF, and with a modified D850 time-of-flight (TOF) reflectron system from Jordan TOF Products, Inc [12]. In the Sciex 4800, a Nd:YAG laser provides 3 to 7 ns long pulses at 355 nm with a repetition rate of up to 200 Hz. The laser pulse energy was 15 µJ and the laser spots had a diameter of 75 µm, resulting in a laser intensity of 340 mJ/cm$^2$. We collected spectra with 900 shots in the mass range from 20 kDa to 500 kDa in linear mode and from 80 Da to 4000 Da in reflectron mode. We measured samples several times in a series of laser spots and we measured samples with varying S1 concentrations. The instrument was calibrated using the mass peaks from a bovine serum albumin (BSA, Merck Life Science NV, Amsterdam, The Netherlands) sample (66.793 kDa).

In our modified Jordan TOF setup, we used a 532 nm laser with 23 ns pulse length (FWHM) at a repetition rate of 10 Hz. Laser pulse fluences were tuned from less than 1 mJ/cm$^2$ to over 500 mJ/cm$^2$ with laser pulse energies ranging from 50 µJ to about 670 µJ/pulse and laser spot sizes from 0.1 to 0.3 mm$^2$. We used a micro-channel plate (MCP) detector as an ion detector in linear mode. The pressure in the sample chamber was < 10$^{-5}$ Torr and we applied differential pumping to maintain pressures below 10$^{-6}$ Torr in the mass spectrometer and near the MCP. The sample was held at a voltage of +1 to +4 kV for acceleration of ions into the linear mass spectrometer.

In Figure 2 we show mass spectra in positive ion polarity from the Sciex 4800 from flat and CNT substrates (both prepared with a drop of 0.2 µg of S1 protein per 1 µl of matrix solution) in the mass range from 20 to 400 kDa. The S1 protein has a mass of 75 kDa. We observe strong mass peaks at 212 kDa, 106 kDa, 71k Da and 53 kDa.  We interpret the peaks as derived from the spike protein S1 trimer at 212 kDa in the q=1 to 4+ charge states, respectively.  The dominant peak at 106 kDa has a full width at half maximum (FWHM) of 10 kDa, broader than the reference peak from the BSA calibration sample, which had a FWHM of 2 kDa at 67 kDa (insert Fig 2, top)).  We observe mass analyte ion intensities that are 50-fold higher from the CNT substrates compared to the flat substrates.

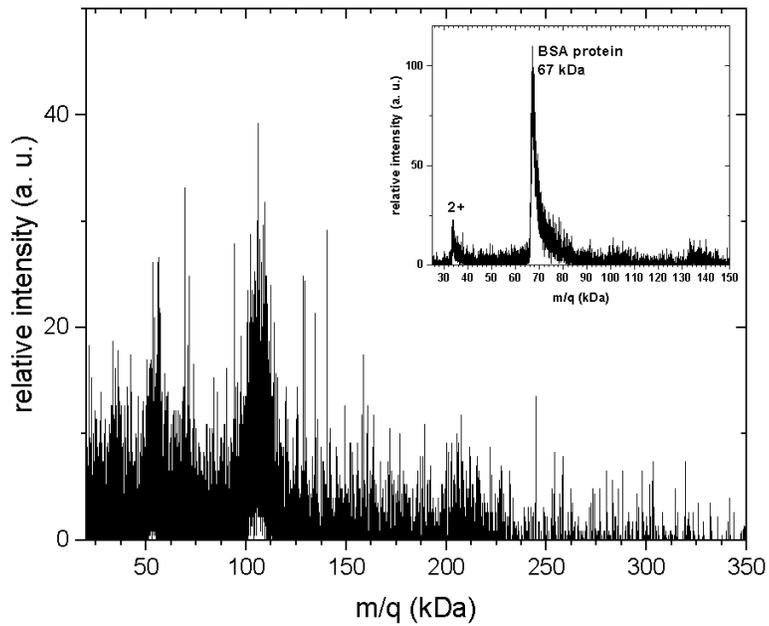

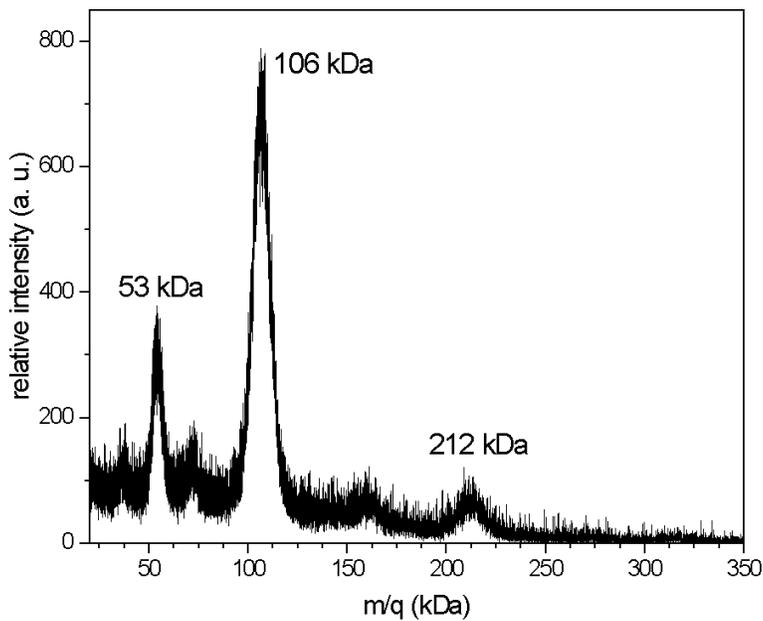

Figure 2: MALDI mass spectra in linear mode from sinapinic acid matrix-spike protein (S1) analyte crystals deposited on a flat silicon sample (top) and deposited on a CNT substrate (bottom). The insert in the top spectrum shows the BSA protein spectrum (from a standard flat sample) used for instrument calibration. Laser exposures were the same in both cases. The spike protein ion yields are enhanced about 50-fold for the CNT sample.

A concern when using CNTs as a substrate in MALDI-MS is potential mass interference from carbon clusters with analyte ions.  In Figure 3 we show mass spectra in reflectron mode in the mass range from 100 to 800 Da for a CNT substrate with matrix solutions (no S1 analyte, top) and with a drop from the matrix-analyte solution (bottom).  These low mass range spectra were collected in reflectron mode with the Sciex 4800 instrument.  The insert in the top spectrum shows a mass spectrum from a CNT sample without matrix or matrix-analyte solution, collected with the Jordan instrument in linear mode.  The CNT sample without matrix deposit shows a series of carbon cluster ions up to 240 Da.  The spectrum from the matrix solution on CNTs shows two dominant peaks from the sinapinic acid matrix at 207 and 224 Da.  We did not observe higher mass carbon cluster ions from the CNT samples.  The observed low mass carbon cluster ions did not interfere with low mass S1 fragment ions.

In the standard MALDI-MS approach, the sinapinic acid matrix facilitates absorption of the laser photons, leading to desorption of intact and fragmented analyte molecules, a fraction of which is also ionized.  The CNT forest can aid absorption of laser energy [13].  This can lead to enhanced desorption and ionization of analyte ions [10].  Laser pulse energy thresholds for efficient ablation and ionization of CNTs are higher than for desorption and ionization of matrix-analyte species, hence mass interferences from CNT based ions is not a concern, especially for the ion mass range above about 200 Da.

We conducted a series of experiments on the trends of substrate, matrix and spike protein ion intensities for a series of laser conditions in order to establish parameters for extraction of optimal signal intensities from a given sample position.  Laser induced desorption processes lead to depletion of matrix-analyte crystals from the sample.  We observe damage spots from laser exposures ex situ by SEM imaging (Figure 1). In Figure 4, we show relative ion intensities in a sequence of 300 laser shots applied to a fixed spot on a CNT sample with matrix-analyte solution.  Analyte ion intensities deplete to ⅓ of their peak intensity after about 60 laser shots.  We observe a threshold laser fluence for desorption of about 20 mJ/cm$^2$ (laser pulse energy of 50 uJ) in the Jorden instrument operated linear mode.  While progressively more matrix crystals are desorbed and the high mass ion yields decreases, yields of low mass carbon ions, which can originate mostly from the CNT substrate, trend higher or remain about constant (insert in Fig 4).

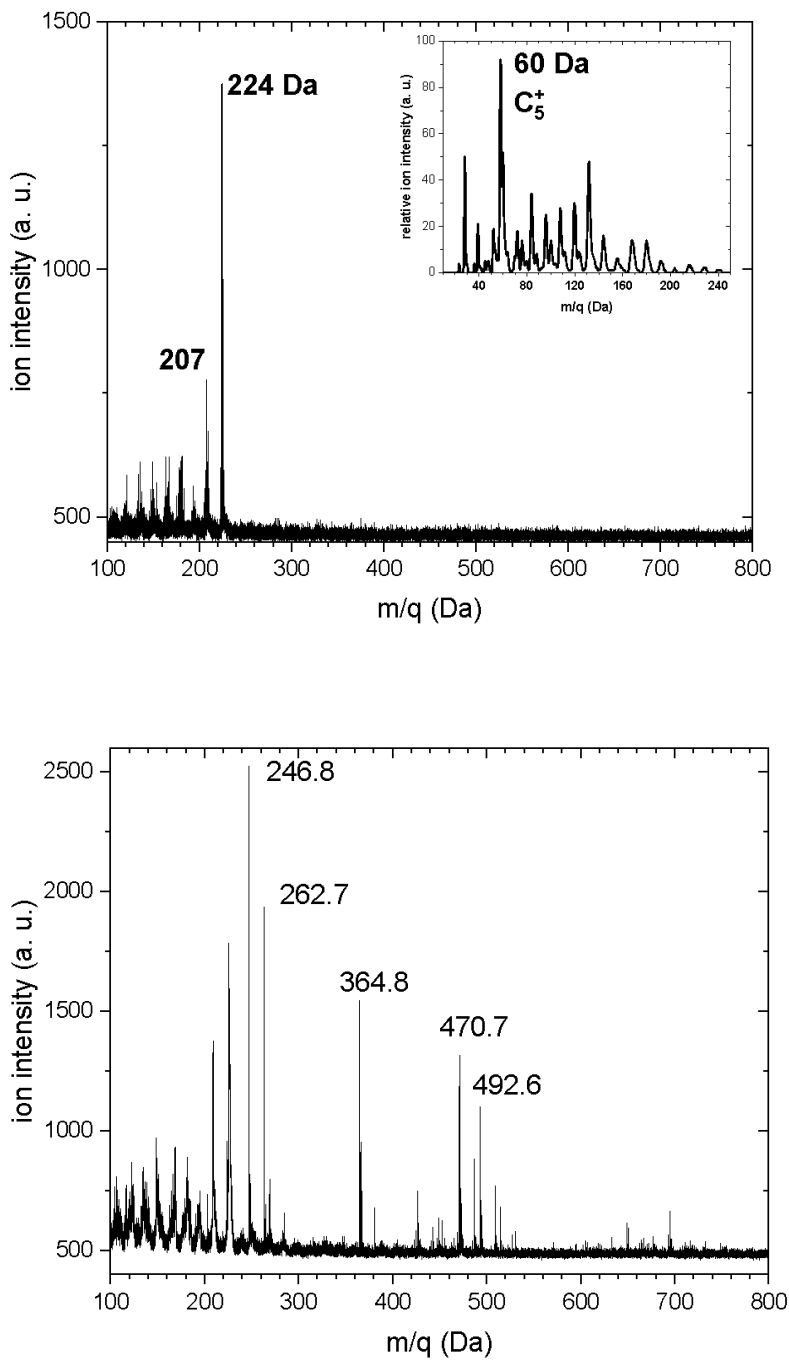

Figure 3: Low mass range ion spectra from MALDI-MS operation in reflectron mode from a matrix-only CNT coated sample (top) and a CNT sample with matrix + analyte solution (bottom). We observe a series of carbon cluster ions up to mass 200 Da, and the main sinapinic acid ion peaks at 207 and 224 Da. The matrix + analyte sample also shows additional peaks in the 250 Da to 700 Da mass range that appear to be S1 fragment ions. The insert in the spectrum on top shows a spectrum from a CNT sample without matrix and without analyte taken in linear mode with the Jordan TOF spectrometer.

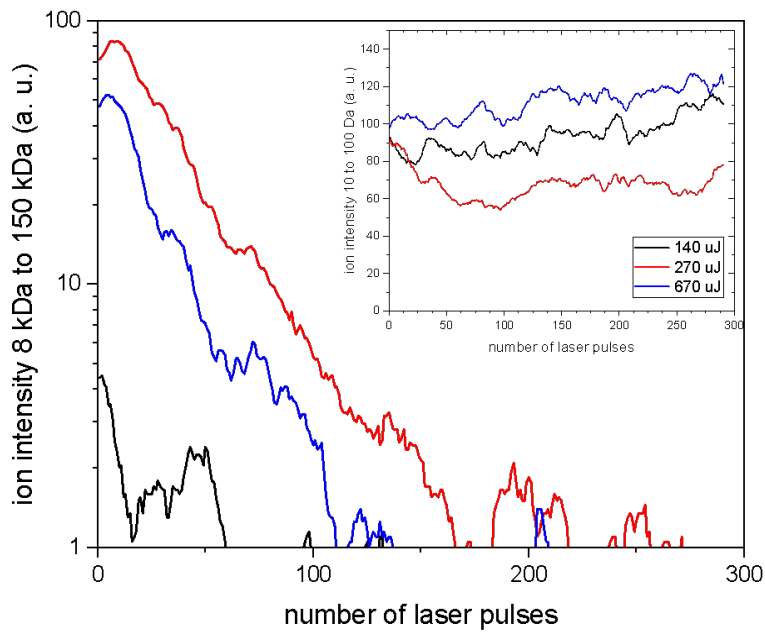

Figure 4: Relative ion intensities in the mass range 8 kDa < M/q < 150 kDa from a CNT sample with matrix + analyte solution as a function of the number of laser pulses for three laser pulse energies (140, 270 and 670 µJ). No ions were detected for a laser pulse energy of 50 µJ. The ion spectra were collected in the Jordan instrument in linear mode. The insert shows relative ion intensities in the mass range 10 Da < M/q < 100 Da as a function of the number of laser pulses.

Spike protein ion yields are very strongly enhanced for CNT substrates compared to flat substrates. We do not have quantitative insights into absolute ion yields or on the ion to neutral fraction in MALDI-MS with CNT substrates or flat substrates. Deposition of drops of matrix-analyte solutions on CNT substrates leads to formation of denser matrix-analyte crystal assemblies compared to flat samples [10]. In addition, absorption of laser light is enhanced by CNT forests [13]. Both effects contribute to the strong increase of analyte ion intensities from CNT samples.

We conducted a series of laser mass spectrometry experiments with CNT substrates but without matrix solutions. Here, spike protein molecules were mixed with artificial saliva. We deposited drops of saliva-analyte solution on CNT substrates and found that the drops spread out over an area of about 1 cm$^2$. MALDI-MS experiments with the Sciex 4800 MALDI instrument did not detect analyte ions from dried drops of salvia-analyte solutions. This underscores the role of the matrix crystals to absorb laser excitations and the role of CNTs to form denser assemblies of matrix-analyte crystals.

We report results from MALDI-MS studies of SARS-Cov-2 spike proteins with nano-structured substrates. We find that samples of matrix-analyte solution prepared with carbon nanotube forests show 50-fold enhanced yields of high mass analyte ions compared to flat substrates. A combined effect of increased densities of matrix-analyte crystals and increased absorption of laser pulse energy with carbon nanotubes leads to strong analyte ion signal increases. Efficient formation and detection of characteristic fingerprint ions can enable detection of pathogens such as SARS-CoV-2 with high sensitivity and specificity. Increased yields of high mass analyte ions from CNT substrates can also be useful for ion injection into MS/MS instruments.  Carbon nanotube substrates and compact laser mass spectrometers can enable low cost testing and environmental monitoring.

In future work, we suggest that it will be interesting to explore ion intensities from intact virus and pathogen samples deposited on CNT forest substrates.  Droplets containing pathogens in matrix solutions could be collected on CNT surfaces, followed by MALDI-MS.  Advances in laser technology make compact short pulse lasers available for integration into mass spectrometers.  Developments in compact particle accelerators e. g. based on MEMS techniques can further support miniaturization of mass spectrometers [14].  We suggest that CNT substrates that collect pathogens and administer efficient absorption of laser energy can increase yields for high mass analyte ions for rapid and reliable identification of pathogens with high sensitivity and in robust instruments that could be widely deployed at relatively low cost.

**Acknowledgments:**

This work was supported by an LDRD at Berkeley Lab, and by the Office of Fusion Energy Sciences and the Office of High Energy Physics in the Office of Science, U.S. Department of Energy, under Contract No. DE-AC02-05CH11231.

**Author contributions:**

TS and AS led the design and execution of the experiments.  TS wrote the article.  KN, PAS, BM, LOH and HK conducted MALDI-MS experiments with the modified Jordon TOF and the Sciex instruments, respectively.  IP conducted SEM analysis.  AJ and RK provided the CNT samples.  All authors contributed to conceptual design of the experiments, data analysis, data interpretation and the editing of the article.